\documentclass[11pt,letterpaper]{article}
\usepackage[utf8]{inputenc}
\usepackage{amsfonts}
\usepackage{amsmath}
\usepackage{braket}
\usepackage{latexsym}
\usepackage{graphicx}
\usepackage{wrapfig}
\usepackage{multicol}
\usepackage{multirow}
\usepackage{fancyhdr}
\usepackage{lastpage}
\usepackage{mathtools}
\usepackage{cite}
\usepackage{abc}

\newcommand{\Z}{\mathbb Z}

\usepackage{lipsum}

\usepackage{color,ifpdf,url}
\title{Computational Complexity of Arranging Music}
\author{
\minipage{0.49\textwidth}
\centering
{\Large William S. Moses}\\
Massachusetts Institute of Technology\\
\small{\href{mailto:wmoses@mit.edu}{wmoses@mit.edu}}
\endminipage
\hfill
\minipage{0.49\textwidth}
\centering
{\Large Erik D. Demaine}\\
Massachusetts Institute of Technology\\
\small{\href{mailto:edemaine@mit.edu}{edemaine@mit.edu}}
\endminipage
}

\newcommand{\term}[2]{\textbf{\it #1}}

\newif\ifabstract
\abstracttrue
\newif\iffull
\ifabstract \fullfalse \else \fulltrue \fi

\usepackage
  [breaklinks,bookmarks,bookmarksnumbered,bookmarksopen,bookmarksopenlevel=2]
  {hyperref}
{\makeatletter \hypersetup{pdftitle={\@title}}}

\advance \topmargin by -\headheight
\advance \topmargin by -\headsep
\evensidemargin \oddsidemargin
{\makeatletter
 \gdef\xxxmark{%
   \expandafter\ifx\csname @mpargs\endcsname\relax 
     \expandafter\ifx\csname @captype\endcsname\relax 
       \marginpar{xxx}
     \else
       xxx 
     \fi
   \else
     xxx 
   \fi}
 \gdef\xxx{\@ifnextchar[\xxx@lab\xxx@nolab}
 \long\gdef\xxx@lab[#1]#2{\textbf{[\xxxmark #2 ---{\sc #1}]}}
 \long\gdef\xxx@nolab#1{\textbf{[\xxxmark #1]}}
}

{\makeatletter \gdef\fps@figure{!htbp}}

\let\realbfseries=\bfseries
\def\bfseries{\realbfseries\boldmath}

\let\epsilon=\varepsilon
\newcommand{\figref}[2][{}]{\hyperref[#2]{\figurename~\ref{#2}#1}} 

\date{}

\begin{document}

\maketitle


\section{Introduction}
Music has long been an interesting subject of analysis for mathematicians and has led to many interesting questions in music theory and other fields. For the most part, computer scientists have looked into applying artificial intelligence to music \cite{AI} and finding algorithms and data structures to solve various problems in music. Prior work on these algorithms often involves computing various properties of music such as the edit distance between two songs \cite{problems} or the optimal fingering \cite{dp}. These problems tend to be solvable in polynomial time using dynamic programming and have various application such as the music identification service Shazam \cite{shazam} or operations on RISM, an online music database \cite{rism}.

This paper takes an additional step in this direction, asking what sorts of problems in music cannot be efficiently computed. Specifically, this paper asks how various constraints affect the computational complexity of arranging music originally written for one set of instruments down to a single instrument. The paper then applies these results to other domains including musical choreography (such as ice skating and ballet) as well as creating levels for rhythm games (such as Rock Band). We prove that all of the problems are NP-complete, meaning that there is no efficient algorithm to solve them (assuming the standard conjecture that P $\neq$ NP).

\subsection{Computational Complexity}
In computer science, algorithms for solving problems are classified by their runtime. For example, a linear scan through a list of size $n$ takes $O(n)$ time. The problems themselves are classified into several \emph{complexity classes} such as P, NP, and EXP. The class P denotes problems for which there exist algorithms that run in polynomial time to solve the problem and EXP denotes problems for which there exist algorithms that run in or faster than exponential time. One of the most studied complexity classes is NP, or nondeterministic polynomial time. Problems in this complexity class are defined to be checkable in polynomial time. The largest open question in computer science right now is whether P $=$ NP --- or in other words whether all problems that can be checked in polynomial time can also be solved in polynomial time. It is widely believed that this is not true and conjectures such as the Exponential Time Hypothesis imply that some NP problems are not solvable in faster than exponential time.

For these complexity classes, we say that a problem is \emph{hard} if it is ``at least as hard as all the other problems in the complexity class.'' Formally, this means that if the hard problem can be solved in polynomial time then all the other problems in the complexity class can be solved in polynomial time. To prove this, we use a \emph{reduction} whereby you take a problem $A$ that you wish to prove hard and show that for all problems $B$ in the complexity class you can create an instance of $A$ which has the same solution to $B$. If there already exists a problem $B$ which is hard for the complexity class, you can prove a problem $A$ hard by simply showing that you can create an instance of $A$ which solves $B$. Additionally, these reductions must take polynomial time to compute. Finally, a problem is complete in a complexity class (e.g.m NP-complete) if the problem is both hard and in the complexity class. The most common example of a problem that is NP-complete is 3SAT, which will be described in Section \ref{3SAT}.

\subsection{Arranging Music}
Often in music, musicians may want to be able to play a piece originally written for one set of instruments for another set of instruments. The resulting song is referred to as an \term{arrangement}. For instance, Bach's Cello Suites are quite commonly arranged for viola. Sometimes musicians may even have to play two parts of one piece (e.g., two violins) on a single instrument (such as a piano). These sorts of modifications are usually successful but often require modification of the original tune to make them fit the specific constraints of the final instrument (e.g., the range of the instrument, number of notes the instrument can play simultaneously, possible fingerings, etc). Formally, we ask whether a set of musical parts $\mathcal{T} = \{\mathcal{T}_0, \mathcal{T}_1, \dots \}$ consisting of musical notes played at specific times, can be arranged or rewritten for a single part, when subject to various constraints defining what sorts of arrangements are permitted.

In this paper, we determine the hardness of arranging music when subject to two sets of constraints, called \emph{specific} and \emph{universal}. Specific constraints represent limitations for the artist playing the arrangement and may be different for different instruments or different artists. Universal constraints represent constraints on arranging music in general. These constraints are used to represent desirable properties whenever arranging any song --- such as requiring that the arrangement be recognizable as the original piece. These universal constraints are unchanged when considering different instrumental limitations. Many of these constraints involve restrictions on \term{chords}, or a set of notes being played simultaneously.

\subsection{Specific Constraints}
Specific constraints considered in this paper come in three varieties: requirements on consonant intervals, limitation in the number of notes in any chord, and limitations in the speed of transitions between notes.

\subsubsection{Consonance Requirement}

First, we consider restrictions on the notes that can appear in chords. This constraint represents that music theory tells us that certain chords naturally sound pleasant (consonant chords) or unpleasant (dissonant chords) to the brain \cite{Cousineau27112012}. Enforcing this constraint allows us to ensure that the resulting arrangement will have only pleasant-sounding chords. Not many songs are completely free of musical dissonance, though most songs tend to have dissonance only sparingly throughout the song.

\subsubsection{Simultaneous Note Limitations}

Second, we consider restrictions on the maximum number of notes that can appear in chords. This constraint represents both instrumental and physical limitations of the artist. For example, a violin has only four strings and thus can play chords with at most four notes. A pianist has ten fingers and can only play ten notes simultaneously. Other instruments such as woodwinds may only be able to play one note at a time. Similarly, less skilled players may not be able to play as many notes.

\subsubsection{Transition Speed Limitations}

Finally, we consider restrictions on the transitions between chords. Specifically, we consider limitations on how quickly notes are permitted to change. This constraint represents that musicians can only play notes so quickly and thus an arrangement requesting them to switch notes more quickly than they can physically play is not useful.

\subsection{Universal Constraints}
In this paper, we consider two universal constraints: disallowing melodies from being split and requiring a certain percentage of the original notes to be in the arrangement.

\subsubsection{No Splitting Melodies}
First, we consider the constraint that each individual part $\mathcal{T}_i$ be included or excluded in entirety. In other words, either all pairs of notes and times associated with part $\mathcal{T}_i$ must be included, or none of them may be. This constraint ensures that melodies are not cut off in the middle. For example if we were arranging ``Twinkle Twinkle Little Star'', we would not want to permit an arrangement that included only the syllables, ``Twin [pause] Twin [pause] Lit [pause] Star'', because such an arrangement is no longer recognizable as the original piece. One may think that this constraint is too restrictive as typically musical parts have a large number of melodies at different times. However, we can first split up each melody into an individual part, and then this constraint becomes valid.

\subsubsection{ Percentage of Notes in Arrangement}
Finally, we consider the constraint that a certain percentage of notes played at any time $t_i$ in the original song are still played in the arrangement. This constraint in particular prevents a valid arrangement from simply not playing any notes, which clearly is not recognizable as the original song.

\subsection{Our Results}
\begin{figure}[!htb]
    \centering
    \begin{minipage}[t]{.54\textwidth}
        \vspace{7mm}
        {
        \begin{tabular}{@{}l | l @{}}
    Problem Variant & Hardness \\ \hline
    $p=0$ & P \\ 
    $p=1$ & P \\ 
    Consonance ($0<p<1$) & NPC \\ 
    Finite Transition Speed ($0<p<1$) & NPC \\ 
    Max $j$-note chord ($0<p \leq \frac{j}{j+2}$) & NPC \\ 
    Max $j$-note chord ($p>\frac{j}{j+1}$) & P \\ 
    Max ($j=1$)-note chord ($p>\frac{1}{3}$) & P \\ 
    \end{tabular}
        }
    \end{minipage}%
    \hfill
    \begin{minipage}[t]{0.44\textwidth}
        \vspace{0pt}
        \includegraphics[width=\linewidth]{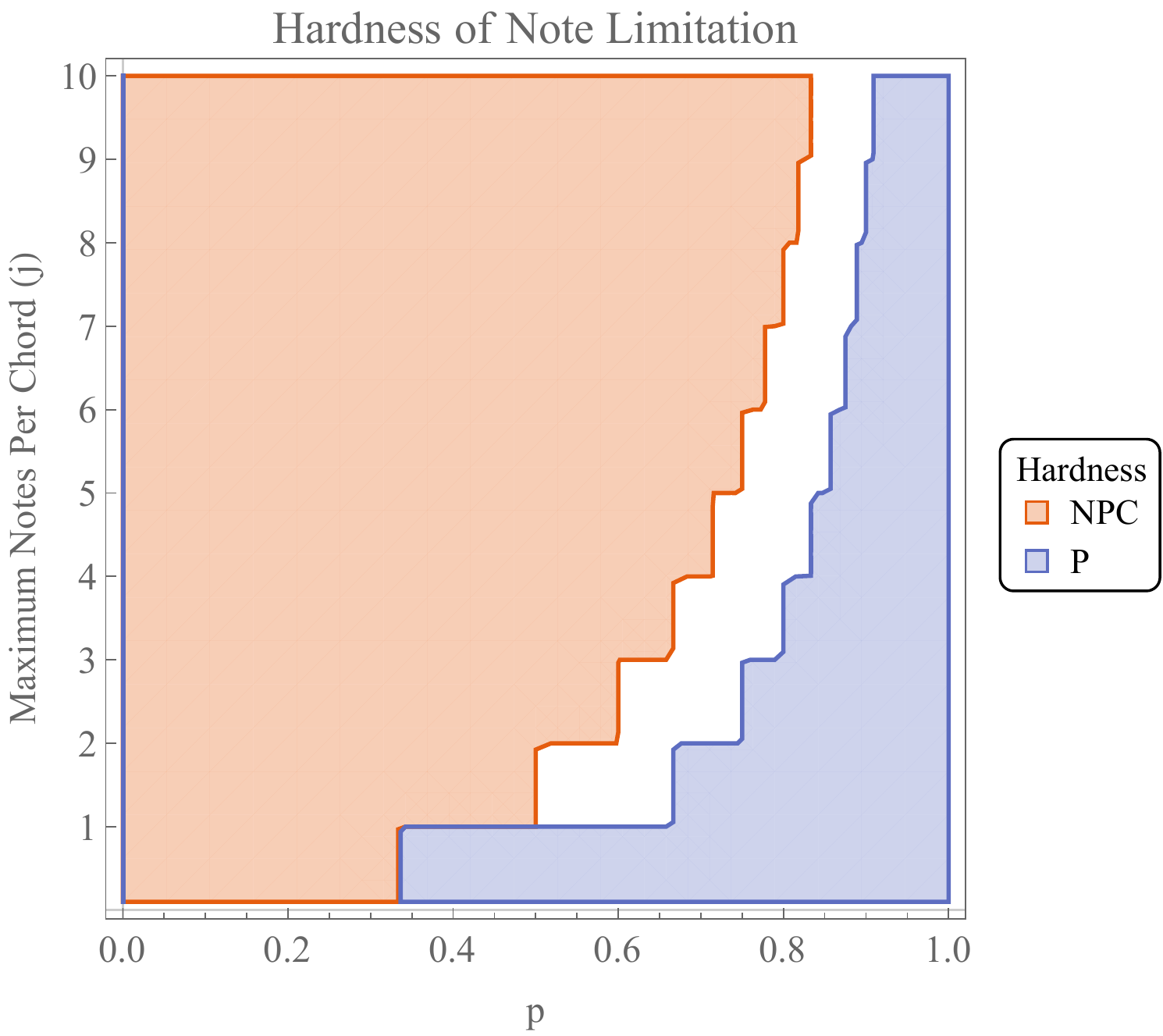}
    \end{minipage}
\caption{A summary of various constraints on the hardness of arranging music. P denotes that the problem is solvable in polynomial time, and NPC denotes NP-completeness. }
    \label{tbl:1}
\end{figure}

\figref{tbl:1} summarizes the computational complexity results we obtain under all combinations of specific and universal constraints. Each specific constraint is considered against all universal constraints. In the majority of combination of these constraints, the problem of arranging music is NP-complete. However, the problem can be solved in polynomial time when subject to specific combinations of these constraints.

\section{Consonant Arrangements}
We begin by considering the problem of creating a musical arrangement that ``sounds good.'' Specifically, we consider the specific constraint that no chords in the arrangement are permitted to be in dissonance (that is, all chords are in consonance). We consider all universal constraints for this problem.

\subsection{Consonance and Dissonance}
\begin{figure}[h!]
  \centering
    \includegraphics[width=0.8\textwidth]{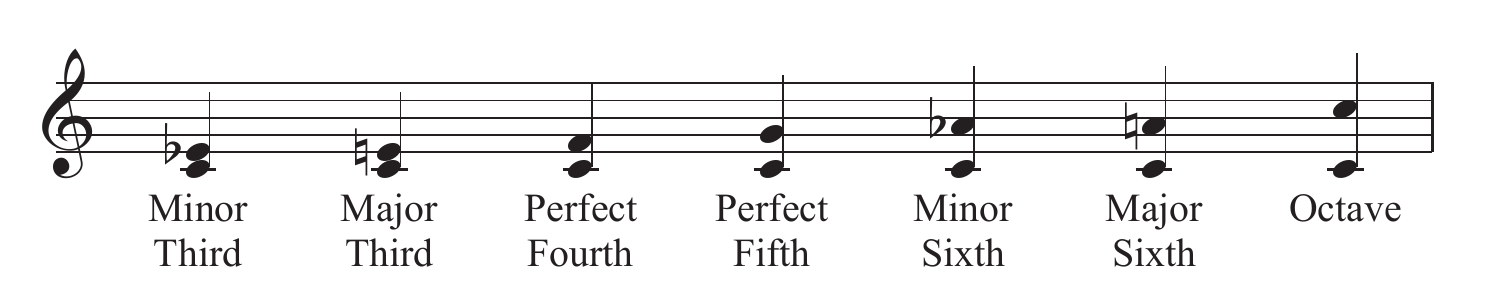}
  \caption{A visualization of consonant intervals.}
    \label{cons}
\end{figure}

First we discuss what entails musical consonance. For now, let us examine only the two-note chords. The number of musical notes (termed half-steps) between the two chords determines whether the two chords are in consonance or dissonance. While there is inherently some subjectivity to which intervals are considered consonant (specifically there is some debate over whether fourths are consonant), we will use the \figref{constab:tab} to define chords as consonant or dissonant. This definition can be visualized by the chords presented in \figref{cons}. Alternate definitions work as well, supposing that the following gadgets are modified to still either be in consonance or dissonance as appropriate.

\begin{figure}[h!]
\centering
Chord Interval Table (in half-steps)\\
\begin{tabular}{ l | c }
  \hline                       
  Consonance & \{0, 3, 4, 5, 7, 8, 9\} + 12$n$, $n\in \Z$ \\
  \hline
  Dissonance & \{1, 2, 6, 10, 11\} + 12$n$, $n\in \Z$\\
  \hline  
\end{tabular}
\caption{A formal definition of which intervals will be considered in consonance.}
\label{constab:tab}
\end{figure}

This idea can then be applied to work with chords of three or more notes by saying that a chord is in dissonance if there is any pairwise dissonance between the nodes of the chord.

The first problem will focus on the hardness of creating an arrangement where all resulting chords are in consonance.

\subsection{Defining the Problem}
This problem asks whether it is possible to make an arrangement of a given song initially written for some number of instruments down to a single instrument that sounds pleasant and remains true to the original song.

Formally, we must satisfy all of the following constraints:
\begin{enumerate}
\item Each individual part is either included or excluded in its entirety.
\item At any given time, at least $p$ notes in the original song need to be played. We'll begin by assuming $p=50\%$, then generalize later.
\item No pair of notes being simultaneously played are in dissonance.
\end{enumerate}

\subsection{Reduction from 3SAT} \label{3SAT}
This problem can be shown to be NP-hard by a reduction from 3SAT. The 3SAT problem asks whether a boolean formula of a certain form can be set to true by setting the variables in the formula appropriately. A 3SAT formula consists of several clauses \textsc{and}ed together, where each clause is an \textsc{or} of three literals (variables or negated variables).  \figref{sat} shows an example.

\begin{figure}[h!]
  \begin{align*}
  (\neg X_1 \vee X_3 \vee X_4) \wedge (X_2 \vee \neg X_3 \vee X_4)
  \end{align*}
  \caption{Example 3SAT having two clauses. One example solution for satisfying this formula is to set all variables to true.}
  \label{sat}
\end{figure}

\begin{figure}[h!]
\centering
\begin{minipage}{.5\textwidth}
  \centering
  \includegraphics[width=0.55\linewidth]{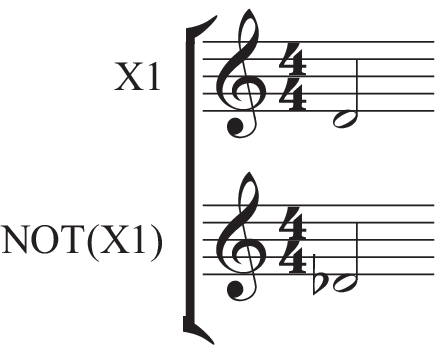}
  \caption{A variable gadget using consonance. Playing both notes in this measure is forbidden as the two notes are a half-step apart which is not a consonant interval.}
  \label{var}
\end{minipage}%
\begin{minipage}{.5\textwidth}
  \centering
  \includegraphics[width=0.55\linewidth]{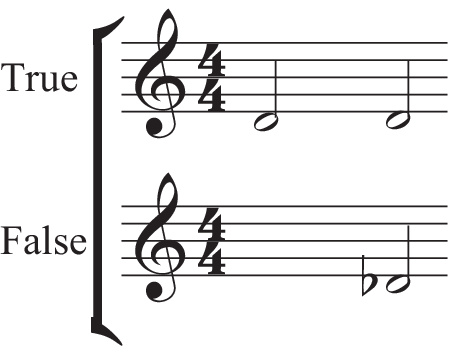}
  \caption{True and false literals respectively for the consonance problem.}
  \label{true}
\end{minipage}
\end{figure}

\subsubsection{Variable Gadget}
We can create a variable gadget by adding two parts per variable, one for true and one for false. At the start of the piece, each variable will have a measure where only the two instruments are playing a note. By having these notes in dissonance with each other, it is not possible to have a valid arrangement if both are played. An example of such literals is shown in \figref{true}.

Combining this with the fact that at least half of all notes at any given time must be played, we are forced to play one and only one of these parts --- thereby forcing the variable to be either true or false.

\subsubsection{True and False Literals}
In order to simply create clause gadgets, we will want to create parts that are guaranteed to be true and false.

We can create a part to represent true (in that all notes in the part must be played) by having a measure where only the true part has notes. Since at any given time at least half of notes must be playing, that note must be played and therefore forces all notes in the part to be played (since parts must be played or not played in their entirety).

We can then create a false literal by creating a variable gadget between a true literal and the literal we intend to create as a false literal. Assuming the variable gadget ensures that only one of the parts can be played, none of the notes in the false literal will be played. An example of such literals is shown in \figref{true}.

\begin{wrapfigure}{r}{0.36\textwidth}
  \centering
  \vspace*{-7mm}
  \includegraphics[width=0.65\linewidth]{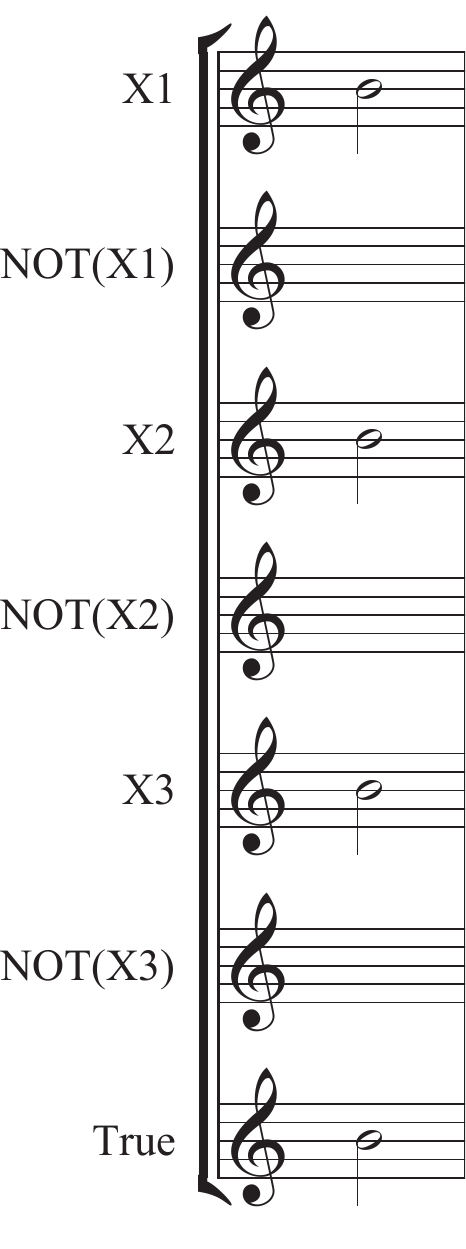}
  \caption{Sample clause for the consonance problem. All notes are identical and thus in consonance.}
\end{wrapfigure}

\subsubsection{Clause Gadget}
We can therefore create a clause gadget by creating a measure where only the true literal and the corresponding variables in the 3SAT clause are playing notes. The requirement that at least half of notes need to be played therefore translates to at least one of the literals being true -- functioning as the requisite OR.

\subsubsection{Generalization in Second Constraint}
We can generalize this proof by permitting $p$ in the constraint requiring ``at any given time at least $p=50\%$ of notes in the original song need to be played'' to take on any rational value between $0$ and $1$. This can be done by padding the gadgets above with appropriate numbers of true and false literals to force the original score to correspond to the selected fraction.

For example, suppose that we selected $p=60\%$ and thus required that at any given time at least three fifths of the notes in the original song need to be played in the arrangement. We cannot use our old clause gadget from above since we would need at least two variable parts to be true for the clause to be satisfied. In the previous clause gadget we had three variable gadgets and one true literal. Setting one variable to true, and the others to false would result in only $50\%$ of the notes being played in that clause. We can can fix this by padding the clause with an additional true literal. Doing so results in three variable parts and two true literals. Thus setting only one variable to true would result in the requisite $60\%$ of notes being played at any time.

More generally, suppose we wanted to build the generalized clause gadget for any given percentage $p$. Suppose we pad the gadget with $t$ true literals and $f$ false literals. If none of the variables are true, then $\frac{t}{t+f+3}$ notes are being played. To ensure that this is a valid clause, this must be strictly less than $p$ since the clause is not satisfied. Likewise, if at least one variable is true, we want to ensure that this clause is satisfied. Hence we also require $\frac{t+1}{t+f+3}\ge p$. Selecting any number of literals $t$ and $f$ that satisfy these two inequalities, we can create a clause gadget for any $p$.

We can show that there always exists a solution to this by first considering how to handle rational values of $p$. First, multiply both the numerator and denominator of $p$ by some large integer. Therefore setting $t+1$ to be the numerator of the number and $t+f+3$ to be the denominator satisfies shows that there exists at least one solution. Likewise, to handle irrational values of $p$, one could first round the number up to a rational number at sufficiently high precision (e.g. round up to nearest $10^{-10}$) and then treat it with the rational number procedure.

\subsubsection{Entire Score}
\begin{figure}[h!]
  \centering
    \includegraphics[width=0.8\textwidth]{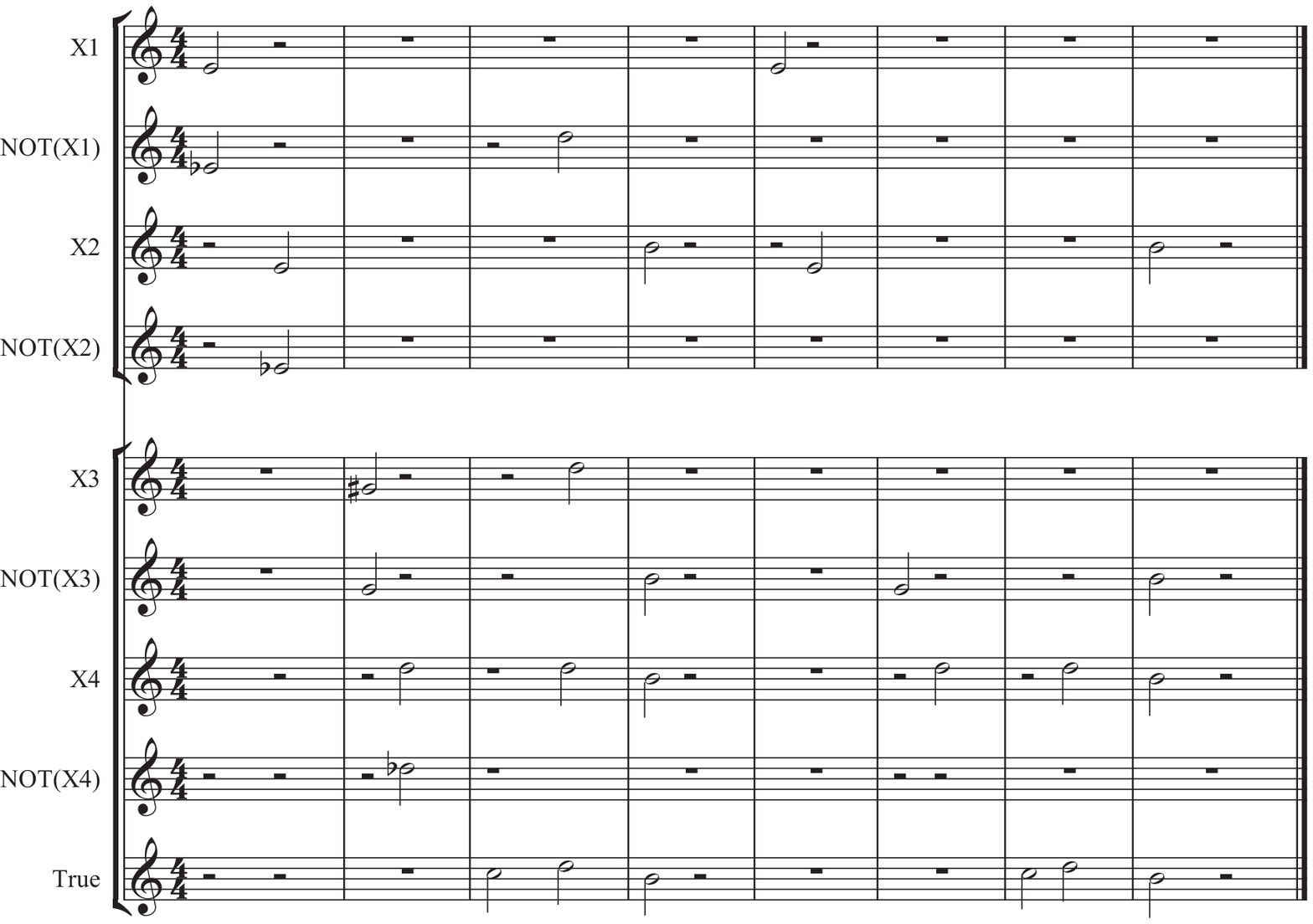}
  \caption{Example reduction for the consonance problem. The first four measures of the piece contain the reduction of the 3SAT. The last four measures of the score contain a valid arrangement and thus satisfying variable assignment to the 3SAT. }
  \label{fullscore}
\end{figure}

A full example of such a reduction is shown in \figref{fullscore}.

In this example the 3SAT formula being reviewed is $(\neg X_1 \vee X_3 \vee X_4) \wedge (X_2 \vee \neg X_3 \vee X_4)$. The first four measures of this song represents the full 3SAT instance. The first two measures are the initialization of the four variable gadgets. The third measure represents the first clause, $(\neg X_1 \vee X_3 \vee X_4)$. The fourth measure represents the second clause, $(X_2 \vee \neg X_3 \vee X_4)$.

The last four measures of the song represents a valid solution to the 3SAT -- namely $X_2,\neg X_3,X_4$. In these measures, there is no dissonance and the $p=50\%$ notes being played at any given time requirement is being satisfied.

\newpage

\section{Limitation in Number of Simultaneous Notes}
This problem will focus on the hardness of creating an arrangement where the number of notes that can be played simultaneously is limited to $j$.

\subsection{Defining the Problem}
Much like the first problem, this problem asks whether it is possible to make an arrangement of a given score for a single instrument that can only play $j$ notes simultaneously and have the arrangement remain true to the original song.

Formally, we must satisfy all of the following constraints:
\begin{enumerate}
\item Each individual part is either included or excluded in its entirety.
\item At any given time, at least $p$ notes in the original song need to be played. We'll begin by assuming $p=50\%$, then generalize later.
\item The number of simultaneous notes in the arrangement is no more than $j$. The number of notes $j \ge 1$.
\end{enumerate}


\subsection{Reduction from 3SAT}
This problem can also be shown to be NP-hard by reduction from 3SAT or X3SAT. X3SAT is a modification to the 3SAT problem, where exactly one literal in any clause has to be true.

\subsubsection{True and False Literals}
We can construct a true literal in the same manner that it was created for the consonance problem as in \figref{true}. The construction of the false literal, however, requires modification since we will want to use it to create the variable gadget and thus cannot use the variable gadget to create the false literal. To create a false literal, we begin by creating $j$ true literals. A false literal can then be created by having a measure with the $j$ true literals and the false literal. Since all true literals must be selected and only $j$ notes can be played, the note in the false part cannot be played, preventing any notes in the false part from being played.

\subsubsection{Variable Gadgets}
Like in the consonance problem, we begin by creating two parts for each variable: one for the true value and one for the false value.

If $j=1$, we can make a variable gadget a similar manner to the consonance problem by having a measure with both true and false versions of the variable. Since we can play at most one note, we are thus forced to play at most one of these parts. Likewise, we must play at least one of these notes since we must play at least $p$ of the notes being played at this time. This holds for any $p>0$.

If $j=2$, we could create a variable gadget by adding one true and and some number of false literals. Since we can only play up to two notes, and we must play the true literal, we can only play at most one of the true or false parts for the variable. In a similar manner to generalizing the clause gadget for the consonance problem, we then pad the variable gadget with sufficient false literals to ensure than the requirement that $p$ notes being played requires us to play the true part and one of the variable parts.

For any $j$, we create a variable gadget by adding $j-1$ true parts and sufficiently many false literals in the same manner for which we built the $j=2$ gadget.

\subsubsection{Clause Gadget}
Let us consider how to construct a clause gadget. 
If $j=1$ and $0<p \le \frac13$, we can construct a clause gadget by having a measure with the three literals in the clause. Since $p$ is between $0$ and $\frac13$, we must play at least one note. Likewise, as a result of the limitation in the number of simultaneous notes, we can only play at most one note. Thus in order for the clause to be playable, one and only one variable can be true. This allows us to reduce from X3SAT to show hardness.

We can extend this technique to higher values of $j$ by padding the clause with $j-1$ true literals. We can show that the problem is NP-hard by the same reduction from X3SAT when $\frac{j-1}{j+2} < p \le \frac{j}{j+2}$. For these values of p, we must select the $j-1$ true literals as well as at least one of the variables. From the limitation in the number of notes, we can only play one of the variables, allowing the reduction from X3SAT to hold.

We can extend this even further by also adding false literals at higher values of $j$. Suppose we padded the clause with $f$ false literals. These allow us to use the same clause gadget as above except now it is valid in the region  $\frac{j-1}{j+2+f} < p \le \frac{j}{j+2+f}$. By combining all possible of the regions shown hard for various numbers of false literals, the problem is thus hard for any $p$, $0< p \le \frac{j}{j+2}$.

\subsubsection{Polynomial Cases}

If $p > \frac{j}{j+1}$, the problem is solvable in polynomial time. This is because for all sets of notes played at the same time, you would either need to play all of the parts containing the notes, or the entire piece could not be satisfied. Consider a moment in the piece when the original song had $N$ possible notes that could be played. Suppose as well that $N \ge j+1$ since otherwise you could successfully play these notes without any problem since you could choose to play any subset of notes. The percentage of notes played at this time is at most $\frac{j}{N}$ since there is the limitation of playing at most $j$ notes. Since $N\ge j+1$, this is at most $\frac{j}{j+1}$. Thus if $p > \frac{j}{j+1}$, this song does not have a valid arrangement. As a result, simply ensuring that at each point in time there are at most $j$ notes that would be needed in the arrangement is sufficient to ensuring that there is a valid arrangement.

When $j=1$ and $ p >\frac{1}{3}$, the problem is solvable in polynomial time by a reduction to 2 coloring. Again suppose you had a moment in the original song that had $N$ playable notes. If $N\ge3$, the piece could not be played since you would not satisfy the requirement that at any time the number of notes being played is greater than $p$. Thus we need only consider $N=1$ and $N=2$. If $N=1$ we must play the note. If $N=2$, this is equivalent to having a choice between either of the two parts with notes -- of which one and only one can be played.

Now, let us create a graph. Take every part in the original piece to be represented by a node in a graph. Then, for every time that two parts have notes at the same time, connect the corresponding graph nodes. Thus checking for a valid two-coloring (where we can consider one color to represent not playing a part and the other color to represent playing a part) is equivalent to the original problem.

\section{Limitations in Transition Speed}

One additional constraint to consider is that there exist limitations on how quickly musicians can change from note to note, or chord to chord. 

\subsection{Defining the Problem}
This problem asks whether it is possible to make an arrangement of a given score for a single instrument that does not require the player to make a transition faster than two beats.

Formally, we must satisfy all of the following constraints:
\begin{enumerate}
\item Each individual part is either included or excluded in its entirety.
\item At any given time, at least $p$ notes in the original song need to be played. We'll begin by assuming $p=50\%$, then generalize later.
\item All notes or chords must be played for at least two beats (a half-note).
\begin{itemize}
\item For example, if Violin I plays a half-note starting at beat 1, and Violin II plays a half-note starting at beat 2, it would be impossible to create an arrangement with both of these notes as it would require playing the lone note of Violin 1 for 1 beat (which is invalid) as well as the chord of both notes for only 1 beat (also invalid), and the lone note of Violin II for 1 beat (invalid as well). 
\item This time is much slower than the actual transition speed of musicians, but the proof can be modified and divide the length of each note and rest by a constant to have a different transition speed.
\end{itemize}
\end{enumerate}

\subsection{Reduction from 3SAT}
This problem also can be shown to be NP-hard by reduction from 3SAT.

\subsubsection{Variable Gadget}
To make this easier to visualize, we will use a time signature that has 8 beats in a measure.
Once again, one can create a variable gadget by adding two parts per variable to represent true and false. To force only one part to be selected, in one measure have the true variable begin playing a note at the third beat and have the false variable begin playing a note at the fourth beat. Additionally, throughout the entire measure the true part will be playing a note. The true literal is necessary otherwise during the 3rd and 4th beats of the measure, the true and false variables will be played along, requiring that both notes be played. The additional true literal satisfies the $\ge50\%$ requirement for these outer notes when the corresponding variable is not selected. The addition of more true literals allows this gadget to work for any value of $p$.

\begin{figure}[h!]
  \centering
    \includegraphics[width=0.4\textwidth]{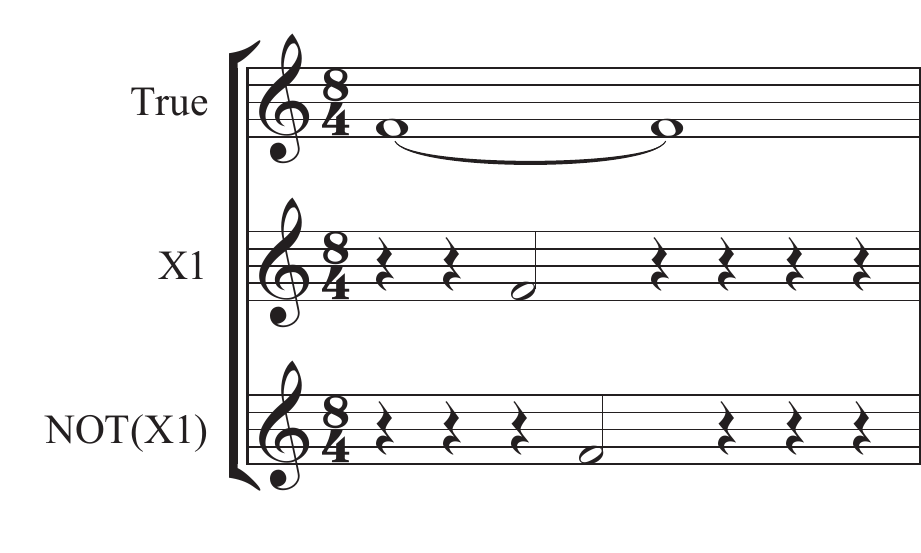}
  \caption{Example variable gadget for limitations in transition speed.}
\end{figure}

\begin{figure}[h!]
  \centering
    \includegraphics[width=0.4\textwidth]{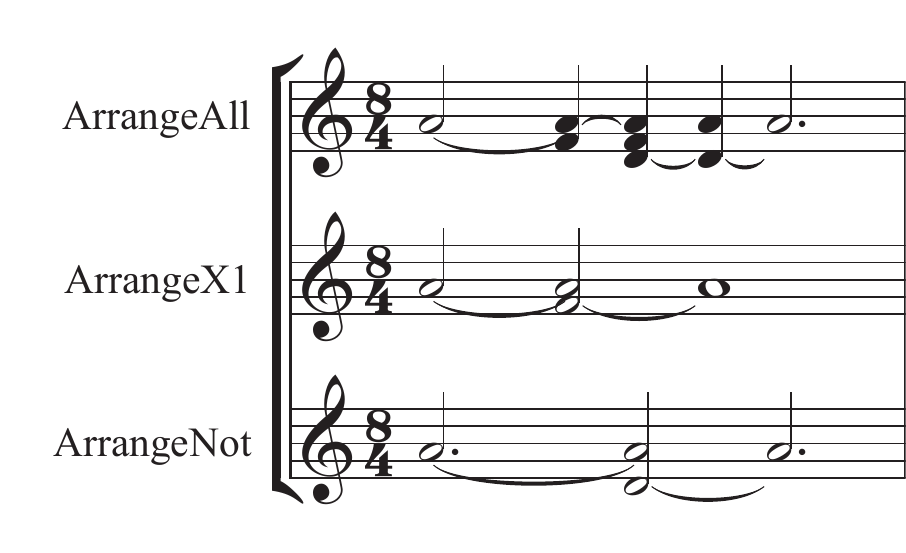}
  \caption{Three arrangements of the 3SAT score, selecting all parts, true and $x_1$, and true and $\neg x_1$, respectively. Selecting the first arrangement is forbidden as it contains three transitions with notes that are one beat long. }
    \label{allspeed}
\end{figure}

To illustrate this, \figref{allspeed} shows three arrangements of the previous 3SAT score. In the first arrangement, you can see a transition between the second and third beats that invalidates the transition. In the second arrangement, only true and $x_1$ where selected. As you can see, there is no note played for less than 2 beats. In the third arrangement, only true and $\neg x_1$ are selected. Once again, no note is played for less than 2 beats.

\subsubsection{True and False Literals}
The true literal can be made in the same way as in previous reductions.

The false literal can be made using a gadget similar to the variable gadget. However, instead of having $x_1$, we use another true literal. This then forces the third part to be false -- thereby creating the false literal.

\subsubsection{Clause Gadget}
The same clause gadget from the consonance problem will work in this scenario.

\section{General Results}
\subsection{Arranging Music is in NP}
Regardless of what set of constraints is considered, the problem of arranging music is in NP. This can be shown by the existence of a polynomial time algorithm which checks that an arrangement of the music is valid or not. This can be done by simply iterating through all the times that notes are played an ensuring that all constraints are being met.

\subsection{Requiring all 100\% of notes played is P}
Regardless of what set of constraints is considered, the problem of arranging music when all notes played in the original song be played in the arrangement is polynomial-time solvable. This is because the only possible arrangement includes all the notes, which simply needs to be be checked by the polynomial-time checking algorithm.

\subsection{Requiring all 0\% of notes played is P}
Regardless of what set of constraints is considered, the problem of arranging music when none of notes played in the original song need to be played in the arrangement is polynomial-time solvable. The solution for this is to simply have an arrangement of no notes and is clearly solvable in polynomial time.

\section{Applications}
These proofs have significant applications to both rhythm gaming and musical choreography.
\subsection{Rhythm Gaming}
The creation of music for video games such as Rock Band or Guitar Hero can be considered direct applications of these proofs. In these scenarios, the original piece of music that one wants to transition to Rock Band is the initial score and the fake guitar is the instrument one wants to arrange for. This application is most well-suited for the problem when the number of notes is limited (since there are only five buttons on the guitar) --- however one could make arguments for the other proofs as well.

This can be extended to rhythm gaming in general where the input device (such as the pad for Dance Dance Revolution or the buttons for Tap Tap Revolution) represents the instrument. As a result one can claim that designing the arrangements for all rhythm gaming is NP-hard.

\subsection{Choreography}
In much the same manner, one can claim that any form of musical choreography is NP-hard. Examples include ballet and ice skating. We extend the definition of an instrument to apply to choreography. In this scenario, various moves would represent the notes on the instrument.

\bibliography{main.bib}{}
\bibliographystyle{plain}

\end{document}